\documentclass[preprint]{aastex}
\usepackage{epsfig}
\shorttitle{PSR J2144$-$3933 ...}
\shortauthors{J. Gil \& D. Mitra}

\begin{document}

 \title{Vacuum gaps in pulsars and PSR J2144$-$3933}

\author{Janusz Gil$^{1}$
         \and
        Dipanjan Mitra$^{2}$
       }


\affil{$^1$J. Kepler Astronomical Center, Lubuska 2, 65-265, 
Zielona G{\'o}ra, Poland\\
 email: jag@astro.ca.wsp.zgora.pl
     \and
$^2$ Raman Research Institute, C. V. Raman Avenue, Bangalore 560080, India\\
Present address: Inter University Center for Astronomy and Astrophysics, Post Bag 4
, Ganeshkhind, Pune 411007, India\\
email: dmitra@iucaa.ernet.in
          }
\def\ni{\noindent}
\def\be{\begin{equation}}
\def\ee{\end{equation}}
\def\stl{\stackrel{<}{\sim}}
\def\stg{\stackrel{>}{\sim}}

\begin{abstract}

In this paper we revisit the  radio pulsar death line problem within
the framework of curvature radiation and/or inverse compton scattering
induced vacuum gap model above neutron star polar caps.  Our special
interest is in the recently detected pulsar PSR J2144$-$3933 with
extremal period 8.5 seconds, which lies far  beyond conventional death
lines. We argue, that  formation of vacuum gaps    requires a
complicated multipolar surface magnetic field, with a strenght $B_s$
much higher than the surface dipolar  component $B_d$, and radii of
curvature ${\cal R}$ much smaller than the neutron star radius
$R=10^6$~cm.  Such a multipolar surface field is also consistent with
death lines including the extremal pulsar PSR J2144$-$3933. Since
vacuum gap  models produce sparks, our paper naturally supports the
spark related models of subpulse drift phenomenon as well  as to the
spark associated models of coherent pulsar  radio emission.
 \end{abstract}


\section{Introduction}

Radio pulsars are believed to turn off when they can no longer produce
electron-positron pairs in strong and curved magnetic fields within a so called 
inner accelerator region just above the polar cap.
The limiting rotational period $P$ at which this occurs depends on the 
magnitude and configuration of the surface magnetic field $B_{s}$. 
Unfortunately, only the dipolar component $B_{d}$ of this field can be 
deduced from the observed spin-down rate. The line on the $B_{d}-P$ 
plane or $\dot{P}-P$ plane corresponding to this critical period is called 
a death line. No radio pulsar should be observed to the right or above of this line i.e. 
with a period longer
than the critical one. Recently, \citet{ymj99} reported on the existence 
of PSR J2144$-$3933 with a period of 8.5 second which is by far the longest 
known.  This pulsar, which is located far beyond all conventional death
lines, seem to challenge existing emission models. As \citet{ymj99}
concluded themselves, under the usual assumptions, this slowly rotating 
pulsar should not be emitting a radio beam.

In this paper we attempt to examine this problem assuming that PSR
J2144$-$3933 generates radio emission just the same way as other pulsars.
Assuming that pulsar's inner accelerator is the Vacuum Gap (VG henceforth) 
driven by Curvature Radiation (CR henceforth) for pair production in 
strong magnetic fields, we 
find that its radio detection is consistent with a
strong and complicated surface magnetic field in pulsars. 
When such fields with a magnitude 
much greater than the measurable dipole component and radius of curvature 
much smaller than the neutron star radius $R=10^6$~cm is assumed, then
one can easily find general death lines which include PSR J2144$-$3933 on their left
side (see Fig.~1). This is consistent with \citet{as79}, who noted that 
if pair creation is essential for coherent radio emission, then pulsars with very
long periods ($\sim 5$ seconds) require more complex surface field than a pure dipole
(see also Arons 2000), where similar conclusions were drawn in a more complete 
treatment,
including the Inverse Compton Scattering (ICS henceforth) as a source of pairs and
frame-dragging effect modifying the electric field near the polar cap surface).
We demonstrate in this paper that formation of
vacuum gap above pulsar polar cap is possible, provided that the surface magnetic field 
is extremely non-dipolar (sunspot-like).

\citet{ymj99} in their discovery paper already noticed that extremely 
curved and strong surface field would solve the problem of death line 
in PSR J2144$-$3933. They were however reluctant to accept that such extreme 
configuration with 
$B_s\gg B_d$ and ${\cal R}\ll R$ can exist. \citet{zhm00} argued that
if the pulsar's inner accelerator is the so-called Space Charge Limited 
Flow (SCLF henceforth) 
with the ICS as the dominating mechanism of pair production then a pure dipolar magnetic field
configuration is sufficient to produce death lines which can explain the presence 
of this pulsar. We however, 
for reasons explained below, will assume that the surface magnetic field configuration 
required by the curvature radiation induced vacuum gap (VG-CR) and/or
inverse compton scattering vacuum gap (VG-ICS) is just what Nature creates in pulsars, which perhaps is not 
the simplest explanation, but nevertheless a possibility, which we intend 
to explore in this paper.

The only strong condition to define pulsar deathlines is the condition for pair production, 
regardless of the detailed 
mechanisms to produce these pairs. Pulsar deathlines are therefore model dependent. 
In history, 
both the VG model and the SCLF model have been developed 
and the corresponding death lines of both models were investigated by 
Ruderman \& Sutherland (1975; RS75 henceforth) and \citet{as79}. 
The original versions of both models assume CR to be the dominating mechanism 
for pair production 
and neglect the influence of the general relativistic frame-dragging effect 
and other sources of pairs. 
Both models were later improved to include the 
frame-dragging effect \citep{mt92} and the ICS \citep{zql97a,zql97b} as an 
alternative mechanism for pair production. \citet{zhm00} 
fully investigated death lines of all different kinds of the models. In \S~2 we argue that both the observational
data and theory of coherent radio emission seem to favour the VG models 
of inner pulsar accelerator. 
We find that the group of pulsars with
drifting subpulses can be naturally explained by the VG-CR model, but to explain 
all the normal pulsars one has to also invoke the
VG-ICS model.

\section{The inner accelerator and observed pulsar radio emission}

As mentioned above, two types of the polar cap acceleration models 
(VG and SCLF respectively) are usually considered. 
In the VG models, the free outflow of charges from the polar cap 
is impeded and the high acceleration potential drop is related to formation of 
an empty gap above the polar cap (Sturrock 1971; RS75; Cheng \& Ruderman 1977, 1980). 
In the SCLF models the charged particles flow freely 
from the polar cap surface and accelerate within a height scale of about one stellar 
radius $R=10^6$~cm due to the potential drop resulting from the curvature of magnetic field lines 
and/or inertia of outstreaming particles \citep{as79,a81}.
Although this potential drop is too weak to explain the entire population 
of existing radio pulsars, \citet{mt92} have shown that the relativistic effect of inertial 
frame dragging generates an additional 
potential drop within about one stellar radius, which is almost two orders 
of magnitude larger than the 
inertial potential drop. In both VG and SCLF models, either the CR or the ICS 
photons should be considered as the sources of pairs.

It is well known that the VG models suffer from the so-called ``binding 
energy problem''
\citep[e.g.][]{hm76,kwmh88}. It was demonstrated that the binding energy of $^{56}$Fe ions is much too 
small to prevent 
thermionic emission from the surface, and thus formation of vacuum gap 
above pulsar polar cap 
was questionable. Thus, the binding energy problem is an important difficulty 
of the VG models 
and in this sense the SCLF models are more natural, unless some further 
assumptions are adopted and justified. Xu et al. (1999; for review see Xu et al. 
2000a) proposed that pulsars showing drifting subpulses, thus requiring some kind
of VG inner accelerator, are bare polar cap strange stars (BPCSS henceforth) 
rather than neutron stars.
This attractive but exotic conjecture and is based on the argument that the 
work function for both electrons and positrons in BPCSS is practically infinite.
In this paper we argue that the binding energy crisis can be solved alternatively 
and more naturally by assuming extremely strong and curved surface magnetic 
field above the NS polar cap. Therefore, we conclude that invoking the BPCSS conjecture is not necessary.

An important feature that distinguishes VG models from SCLF models is that 
the former produce sparks, as originally delineated by RS75. Each spark generates 
a plasma cloud and interaction of adjacent clouds may have something
to do with the generation of coherent radio emission (see below). Due to 
$\mathbf{E}\times\mathbf{B}$
drift within the gap, the subpulse associated sparks are expected to rotate 
slowly around the magnetic pole,
at a rate which can be tested observationaly in pulsars showing drifting 
subpulses in their single pulse emission.
Recently, \citet{dr99} analysed drifting subpulses in PSR B0943$+$10 with a newly 
developed technique called ``cartographic transform'', based on the sophisticated
fluctuation spectra analysis.
They obtained a clear map of the polar cap with 20 sparks rotating around 
the pole at the rate consistent with the RS75 model, with each spark completing one full
rotation in 37 pulsar periods. 
It is this consistency which tells us that the RS75-type VG model is 
realized in nature, at least in pulsars showing
periodicies in their fluctuation spectra that can be associated with 
rotating sparks. It is important to emphasize
that it is not just a handful a pulsars
with clearly drifting subpulses. \citet{r86} demonstrated, that such 
periodicities in the range 2-15 pulsar periods are common among 
the so called conal profile components, while core components either 
do not show any evidence of periodicities or they are 
much weaker (in the range 15-50 pulsar periods). In Fig.~1 we marked by 
crossed circles 41 pulsars showing
periodic subpulse modulations, following \citet[][Tables~2 and 3]{r86}.

\citet{gs00} reanalysed the case of PSR B0943$+$10 and three other pulsars 
with a clear subpulse drift (see Table 1 in this paper), within their
modified RS75 model. They confirmed consistency of the observed 
periodicities with their modified RS75 VG model. The fundamental drift
periodicity can be expressed by $P_3\approx P\cdot a^2/N$, 
where $a=r_p/h$ (the complexity parameter) is the ratio of
the canonical polar cap radius to the gap heigh, $P$ is the pulsar 
period and $N$ is the number of sparks circulating around
the pole. For PSR B0943+10 $a\approx 6$, $N=20$ and therefore $P_3\approx 1.8P$, 
as observed \citep[e.g.][]{dr99}. For small
values of $a$, $P_3<2P$ and the subpulse drift is hard to be 
detected. For medium values of $a$, the $P_3$
values are in the range $(2-15)P$, and for very large $a\gg 10$ 
(core single pulsars - see Gil \& Sendyk 2000), the $P_3$ 
periods largely exceed $15P$.

The fundamental problem of pulsar studies is that there is no 
consensus at all about the radio emission mechanism 
except that pair production is an essential condition and
that the radiation mechanism must be coherent. Yet another widely accepted 
constraint is that the emission region
is located close to the neutron star, at altitudes about few percent of 
the light cylinder radius $R_{LC}=cP/2\pi$ \citep{c92,kg97}. However, in the millisecond 
pulsars it can reach even $\sim 30\%$ of $R_{LC}$ \citep{kg98}.

After more then 30 years of intensive research, only few succesful, 
self-consistent models of generation of coherent pulsar radio emission can 
be found in the literature. Historically, the first one (called the Georgian model)
was proposed by \citet{kmmu87,kmm92}. In this model, based on the SCLF inner 
acceleration scenario, the radio emission
is generated by a maser relativistic plasma radiation.
This model requires relatively low magnetic field, thus high emission altitudes (larger 
than 30\% of $R_{LC}$), and therefore its possible applicability is 
restricted to millisecond pulsars.

Qiao and collaborators \citep[1998; for review see][]{xlhq00} proposed 
a coherent ICS model of pulsar radio emission. This model
is based on the sparking VG scenario \citep{xlhq00}. The low frequency 
electromagnetic wave associated with
development and decay of a spark are scattered on bunches of 
outstreaming particles. The coherent ICS model
seems to have some observational difficulties. As \citet{xlhq00} argued, it 
can reproduce the core pulsar beam
and two conal (inner and outer) beams. However, present day observational data 
suggest that pulsar beams may
consist of up to three or even four nested cones \citep{md99,gs00}, which cannot 
be explained within the ICS model, at least by its simplest published version 
(including a double-peaked pair distribution function will probably result in more cones;
Zhang 2000, privat information).  
The more serious problem is the observed pulsar polarization. The coherent 
ICS model predicts that the individual emission corresponding
to the core beam should be circularly polarized, while the conal beams 
should demonstrate no significant circular polarization
\citep{xlhq00}. However, available polarimetric observations of single pulses 
show that subpulses in conal pulsars are 
clearly circularly polarized, with sense reversals near the subpulse 
peak, consistent with the model of coherent curvature 
radiation \citep{g92,gan99}. This property, if confirmed in larger sample
of pulsars, might be a big challenge for the 
coherent ICS model of pulsar radio emission.
The same applies to the maser relativistic plasma radiation (Georgian model).

Recently, a new idea has been developed, according to which the observed 
pulsar radio emission is a coherent
curvature radiation emitted by charged solitons associated with sparks 
operating in the inner VG accelerator \citep{am98,gs00,mgp00}. The model is 
entirely self-consistent and determines pulsar characteristics by two
basic observables, $P$ and $\dot{P}$. The present version of the soliton 
model explores the VG-CR model, although
the VG-ICS alternative is not excluded. 

Pulsars with drifting subpulses, which require existence of sparks, are
distributed more or less uniformly over the bulk of typical pulsars on the $P-\dot{P}$ diagram (Fig. 1).
On the other hand, radio emission of typical pulsars originates at low altitudes, which favours radiation
models based on the sparking scenario. 
 It is thus clear from the above discussion that the VG-CR and/or VG-ICS models seem to be 
preferred in typical pulsars, both from the observational and
theoretical point of view. We therefore concentrate in this paper on this class of
inner accelerator models and we plan to
give a full treatment, including the SCLF along 
with the frame dragging effect (which perhaps can be applied to millisecond 
pulsars) in a separate paper. 

\section{Vacuum gap formation in superstrong magnetic field}

We will argue in this paper that the actual surface magnetic field
is extremaly strong and curved with $B_s\gg B_d$ and ${\cal R}\ll R$,
where $B_d=6.4\cdot 10^{19}(P\cdot\dot{P})^{1/2}$~G is the global surface dipole component
\citep[e.g.][]{zh00b},
${\cal R}$ is the radius of curvature of surface field lines and $R=10^6$~cm is the NS radius. 
Thus, for a convenience of further considerations we present in this section new results 
concerning gap formation and
death lines in superstrong magnetic fields $B_s>0.1 B_q$, where $B_q=m^2c^3/e\hbar=4.4\times 10^{13}$~G
is the critical magnetic field strenght at which the electron gyrofrequency $\hbar\omega_c=\hbar eB/mc$
is equal to its rest mass. Above this strengh the photon splitting phenomenon may inhibit pair formation
process \citep{bh98}. More exactly, the photon splitting becomes effective above the critical field strenght 
$B_c \simeq(5.7\times 10^{13}~G)P^{2/5}$ \citep[e.g.][]{zh00b}.

In the superstrong magnetic field $B>0.1B_q\approx 5\cdot 10^{12}$~G, the 
high energy photons with frequency
$\omega$ will produce electron positron pairs at or near the kinematic 
threshold \citep[e.g][]{dh83}
\be
\hbar\omega=2mc^2/\sin\theta ,
\label{a1}
 \ee
where $\sin\theta=l_{ph}/{\cal R}$, $l_{ph}$ is the photon mean free path 
for pair formation and
${\cal R}={\cal R}_6\cdot 10^6$~cm the radius of curvature of magnetic field 
lines within the gap. The typical photon energy is
\be
\hbar\omega=(3/2)\hbar\gamma^3c/{\cal R}
\label{a2}
\ee
in case of curvature radiation (e.g. RS75), and
\be
\hbar\omega=2\gamma\hbar eB/mc
\label{a3}
\ee
in the case of the resonant inverse compton scattering \citep[e.g.][]{zql97b}. 
Here $\gamma$ in the typical Lorentz
factor, $\hbar$ is the Planck constant, $m$ is the electron mass, $e$ is the 
electron charge and $c$ is the
speed of light. It is worth to emphasize that the near threshold gap parameters have not been studied before.

\noindent {\bf Curvature radiation induced near threshold vacuum gaps}

Let us consider the CR photons as sources of pairs first. The gap height $h$ is determined by the 
condition $h=l_{ph}$. The Lorentz
factors $\gamma$ can be calculated from the potential drop across the gap
\be
\Delta V=(2\pi/cP)\cdot B_s\cdot h^2
\label{a4}
\ee
as
\be
\gamma=\frac{e\Delta V}{mc^2} .
\label{a5}
\ee
If we express the surface magnetic field as $B_s=b\cdot B_d=2\cdot b\cdot(P\cdot\dot{P}_{-15})^{1/2}$
then equations (\ref{a1}), (\ref{a2}),(\ref{a4}) and (\ref{a5}) give the gap height in the form
\be
h=3\cdot 10^3{\cal R}_6^{2/7}b^{-3/7}P^{1/7}\dot{P}_{-15}^{-1/7} ~~{\rm cm},
\label{a6}
\ee
where $0.1B_q<b<B_q/B_d$ and ${\cal R}_6\stl 1$ ($B_q=4.4\cdot 10^{13}$~G).
This can be compared with the asymptotic case \citep{e66} valid for $B\leq 0.1B_q$
\be
h_{RS}=3.5\cdot 10^3{\cal R}_6^{2/7}b^{-4/7}P^{-1/7}\dot{P}_{-15}^{-2/7} ~~{\rm cm} ,
\label{a7}
\ee
where $b\stg 1$ and ${\cal R}_6\sim 1$ (RS75).

To obtain the critical lines for the VG formation, we use 
the condition $T_s/T_i\leq 1$, where the ion critical thermonic temperature above which
$^{56}$Fe ions will not be bound
\be
T_i=10^7(B_s/10^{14}{\rm G})^{0.73}\approx 6\cdot 10^5\cdot b^{0.73}(P\cdot\dot{P}_{-15})^{0.36}~{\rm K}
\label{8}
\ee
\citep{as91,um95,zh00b} and the actual surface temperature
\be
T_s=(\kappa\cdot{\cal F})^{1/4}\left(\frac{e\cdot\Delta V\cdot\dot{N}}{\sigma\cdot\pi\cdot r_p^2}\right)^{1/4} ,
\label{a8}
\ee
where $\dot{N}=\pi r_p^2\cdot B_s/(eP)$ is the particle flux through the polar cap with radius 
$r_p=1.4\cdot 10^4\cdot b^{-0.5}P^{-0.5}$~cm, $\Delta V$ is
expressed by equation (\ref{a4}) and reduction parameters $\kappa\approx{\cal F}\stl 1$ are
described in \S 4. The family of 
critical lines for VG-CR
formation in the superstrong and extremely curved surface magnetic field has therefore a form
\be
\dot{P}_{-15}=2.7\cdot 10^3(\kappa\cdot{\cal F})^{1.15}{\cal R}_6^{0.64}\cdot b^{-2}P^{-2.3} .
\label{a9}
\ee
with actual lines depending on values of parameters $b$ and ${\cal R}_6$ (as well as $\kappa$ and ${\cal F}$).
The curvature radiation induced vacuum gap can form in pulsars 
lying above these lines. The parameter
space for the VG-CR inner accelerator is approximately determined by the
two extremal lines corresponding to 
$(\kappa\cdot{\cal F})^{1.15}\cdot{\cal R}_6^{0.64}\approx 0.1$ and $b\approx B_q/B_d$ for the lower line, and
$(\kappa\cdot{\cal F})^{1.15}\cdot{\cal R}_6^{0.64}\approx 1$ and $b\approx 0.1B_q/B_d$ for the upper line.

The pulsar death line can be defined (e.g. R75) by the condition 
that the actual potential drop accross the
gap accelerator $\Delta V=(2\pi/cP)\cdot B\cdot h^2$ (eq. [\ref{a4}]) 
required to produce enough pairs per primary
to screen out the gap electric field is larger than the maximum potential 
drop $\Delta V_{max}=(2\pi/cP)\cdot b\cdot h_{max}^2$ available from the 
pulsar, in which case no secondary pairs will be produced. Since
$h_{max}=r_p/\sqrt{2}$ (RS75), where $r_p=1.4\cdot b^{-0.5}\cdot 10^4P^{-0.5}$~cm 
is the polar 
cap radius, then using equation (\ref{a6}) for the gap height $h=h_{max}$ we obtain 
family of death line equations for the VG-CR in the form
\be
\dot{P}_{-15}=2.4\cdot 10^{-4}{\cal R}_6^2b^{0.5}P^{4.5} .
\label{a10}
\ee
All pulsars driven by the VG-CR inner accelerator should lie to the left 
of these lines on the $P-\dot{P}$ diagram,
if the surface magnetic field is stronger than about $5\cdot 10^{12}$~G. 
The extremal line is determined by maximum value
of $b\sim B_q/B_d$ and minimum value of ${\cal R}_6\sim 0.1$.

\noindent{\bf Inverse Compton Scattering induced near threshold vacuum gaps}
 
In this case we have to consider the mean free path $l_e$ of electron/positron 
moving with Lorentz factor $\gamma$
to emit one photon with energy expressed by equation (\ref{a3}), since 
$l_e\sim 0.00276\gamma^2B_{12}^{-1}T_6^{-1}\sim l_{ph}$ 
\citep{zhm00} while $l_e\ll l_{ph}$ in the CR case. Here 
$B_{12}=2\cdot b(P\cdot\dot{P}_{-15})^{1/2}$ and $T_6$ is the surface
temperature in units of million K. Thus, assuming that $l_{ph}\sim l_e$ 
we obtain from 
equations (\ref{a1}) and (\ref{a3}) the typical Lorentz factors 
$\gamma=2.2\cdot 10^7\cdot h^{-1}{\cal R}_6\cdot b^{-1}(P\cdot\dot{P}_{-15})^{-0.5}$. 
The surface temperature obtained from equation (\ref{a8}) is $T_6=T/10^6{\rm K}=
0.06\cdot b^{0.5}\cdot h^{0.5}\dot{P}^{0.25}_{-15}P^{-0.25}$ and the electron 
mean free path $l_e\approx 10^{13}{\cal R}_6^2b^{-3.5}h^{-2.5}P^{-1.25}\dot{P}_{-15}^{-1.75}$. 
Now setting $h=l_{ph}=l_e$ \citep{zhm00} we can find the ICS induced
near threshold ($B>5\cdot 10^{12}$~G) gap height
\be
h=5\cdot 10^3{\cal R}_6^{0.57}b^{-1}P^{-0.36}\dot{P}_{-15}^{-0.5} ~~{\rm cm} ,
\label{a11}
\ee
where $b\gg 1$ and ${\cal R}_6\ll 1$. This $h$ can be compared with the 
asymptotic case ($B<5\cdot 10^{12}$~G) given by 
\citet{zhm00}
\be
h=8.8\cdot 10^3{\cal R}_6^{0.57}b^{-1.57}P^{-0.64}\dot{P}_{-15}^{-0.79} ~~{\rm cm},
\label{a12}
\ee
where $b\stg 1$ and ${\cal R}\sim 1$.
 
Now, getting back to the surface temperature expressed in terms of $h$ we obtain
\be
T_s=4\cdot 10^6(\kappa\cdot{\cal F})^{0.25}{\cal R}_6^{0.28}P^{-0.43}~{\rm K} .
\label{a13}
\ee
The gap condition $T_s/T_i=7\cdot{\cal R}_6^{0.28}b^{-0.73}P^{-0.79}\dot{P}_{-15}^{-0.36}\leq 1$ gives the
family of critical lines
\be
\dot{P}_{-15}=2\cdot 10^2\cdot(\kappa\cdot{\cal F})^{0.7}{\cal R}_6^{0.8}b^{-2}P^{-2.2} ~,
\label{a14}
\ee
with actual lines depending on values of parameters $b$ and ${\cal R}_6$ (as well as on $\kappa$ and ${\cal F}$).
The ICS induced vacuum gap in strong magnetic field $B>5\cdot 10^{12}$~G 
can form in pulsars lying 
above these lines, as long as $B<B_q\sim 5\cdot 10^{13}$~G.

To obtain a death-line for the near threshold VG-ICS we need to equate 
the gap height expressed by equation (\ref{a11})
to $r_p/\sqrt{2}$, which leads to family of lines
\be
\dot{P}_{-15}=0.25\cdot{\cal R}_6^{1.14}b^{-1}P^{0.28} .
\label{a15}
\ee
All pulsars driven by strong field ($B>5\cdot 10^{12}$~G) VG-ICS inner accelerator should lie above these
lines on the $P-\dot{P}$ diagram.

\section{Binding energy problem and vacuum gaps in pulsars}

As discussed above, recent analysis of drifting subpulses in 
pulsars \citep{dr99,vj99,gs00} strongly suggests that
sparks rotating around the magnetic pole in vacuum gap do exist, at least in some pulsars.
Such a scenario was first proposed by RS75, but then it was criticised 
due to the so called ``binding energy
problem''. However, the binding energy calculations were made under 
the assumption of a global dipolar magnetic
field above the polar cap. Here we discuss an influence of 
strong multipolar components dominating the surface
field on the formation of vacuum gap above the  neutron star polar caps.

Let us begin with standard approach based on the classical work of RS75.
Within the vacuum gap, the high potential drop discharges via a 
number of isolated sparks. This number
is roughly equal to $a^2$, where $a=r_p/h$ is the so called 
``complexity parameter'', $r_p=10^4P^{-1/2}$ cm is
the canonical polar cap radius and $h$ is the polar gap height, which 
is also the spark characteristic dimension \citep{gs00}. The effective 
surface area beneath $a^2$ sparks in strong surface magnetic field $B_s=b\cdot B_d$
is $A_{eff}\approx a^2\pi(h/2)^2=\pi{10^8}/(4b\cdot P)=A_{GJ}/(4\cdot b)$, 
where $A_{GJ}$ is the canonical area
of the Goldreich-Julian (1969) polar cap with dipolar field. 
The maximum back-flow current of electrons (positrons)
heating the polar cap beneath sparks is $I_{max}=e\cdot\dot{N}_{max}/(4b)$, 
where $\dot{N}_{max}$ is the maximum
available flux density corresponding to final stage of development 
of the spark's plasma (eq. [1] in RS75). At this
stage the potential drop beneath the spark is $\Delta V={\cal F}\cdot V_{RS}$, 
where $V_{RS}$ is the vacuum gap
potential drop (see RS75 - their eq. [23]) and ${\cal F}<1$ is the reduction 
factor describing the voltage discharge
at final stages of the spark developing process. The parameter ${\cal F}=1$ 
only in the empty gap but when sparking
avalanche develops, its value should drop significantly below unity (vacuum 
value). In fact, when a spark ignites,
its first shower deposits very few charges at the gap boundaries, and the
voltage across the gap is almost the maximum vacuum value. But few microseconds 
later the shower flux density
reaches maximum value and the gap voltage beneath the spark is reduced significantly, 
to values inhibiting further effective pair production \citep{gs00}. At this 
stage heating is much more effective than at the very beginning of
the spark discharge. Moreover, this effective heating stage is not reached at the same 
time in all adjacent sparks. We will therefore
introduce below the ``heating efficiency'' factor $\kappa<1$. The energy 
deposited by cascading charges onto
the polar cap surface will diffuse
into deeper layers of the crust, and diffuse out in a later time 
\citep[e.g.][]{ec89}. Some heat may not be transfered
back to be reemitted from the surface. We will 
make all calculations for ${\cal F}\approx\kappa\approx 1$
and disscus a possibility that both these 
parameters are slightly lower than unity.

The thermal X-ray flux from the polar cap populated with $a^2$ sparks is
$L_x=\kappa\cdot\Delta V\cdot I_{max}=\kappa {\cal F}\dot{E}_x/(4b)$,
where $\dot{E}_x=10^{30}b^{6/7}B_{12}^{6/7}{\cal R}_6^{4/7}P^{-15/7}$~erg/s
is the upper bound for the energy flux carried by relativistic positrons into 
the magnetosphere above the gap
(see RS75 - their eq. [26]). Since $L_x=A_{eff}\cdot\sigma\cdot T_s^4$, where
$\sigma=5.7\times 10^{-5}{\rm cm}^{-2}{\rm K}^{-4}$~erg/s, we obtain an 
estimate of the actual surface temperature $T_s\approx 3\cdot(\kappa\cdot{\cal F})^{0.25}
\cdot 10^6b^{0.21}{\cal R}_6^{0.14}\dot{P}_{-15}^{-0.43}P^{0.14}$~K.
The iron critical temperature $T_i$ above which the $^{56}$Fe ions will not be bound is described by
equation (\ref{8}).
Using the ratio of the surface $T_s$ and ion $T_i$ temperatures
\be
\frac{T_s}{T_i}\approx 5\cdot(\kappa\cdot{\cal F})^{0.25}b^{-0.52}{\cal R}^{0.14}_{6}P^{-0.22}\dot{P}_{-15}^{-0.79} ,
\label{tsti}
\ee
we can write a condition for the formation of vacuum gap $T_s/T_i\stl 1$ in the form  
\be
\dot{P}_{-15}>7.7\cdot(\kappa\cdot{\cal F})^{0.32}b^{-0.66}{\cal R}_6^{0.18}P^{-0.28} .
\label{dotp}
\ee 
If we now use $\kappa\cdot{\cal F}=1$, ${\cal R}_6=1$ and relatively high $b=5$, 
then the critical line above which the VG-CR can form is $\dot{P}_{-15}>2\cdot P^{-0.28}$.
As one can see, this line marked in Fig.~1 as the line (1) leaves almost half of 
the known pulsars below it.

In an attempt to explain pulsars below the line (1), let us now consider 
superstrong surface magnetic field $B_s\gg B_d$. In such
a strong field, the pairs are created near the kinematic threshold 
$(\hbar\omega/2mc^2)(h/{\cal R})\approx 1$
\citep{dh83}.
A general description of pair creation processes in such a strong surface 
magnetic field for both CR and ICS induced gaps
is presented in the \S~3. The near threshold critical lines for the VG-CR 
formation are described by equation (\ref{a9}).
To include possibly large number of pulsars lying below the low $B_s$ 
assymptotic line (1), we will assume $b=100$ (which requires $B_d\stl 5\cdot 10^{11}$~G
so that $B_s$ does not exceed $B_q$) and 
${\cal R}_6^{0.64}(k\cdot{\cal F})^{1.15}=0.15$. This gives an extremal critical
line $\dot{P}_{-15}=0.04\cdot P^{-2.3}$, which is presented as the line (2) 
in Fig.~1. 

Most pulsars with drifting subpulses (crossed circles in Fig.~1) can be then explained 
by the critical lines (1) and/or (2), corresponding to the curvature radiation 
dominated vacuum gaps. Table 1 presents
four pulsars modelled carefuly by \citet{gs00} within their modified RS75 drift model. If 
we assume ${\cal R}_6=0.3$ and $b=10$ in equation (\ref{tsti}), then $T_s/T_i\leq 1$, so 
the VG-CR can form in these pulsars, although relatively strong surface magnetic 
field $B_s\sim 10B_d$ is required. Such strong magnetic field is consistent with the 
recently discovered longest period pulsar PSR J2144$-$3933, which we discuss in 
the next section. 
In the summary session we discuss observational constraints for the strenght
of the surface magnetic multipole components and argue that strong multipole 
fields with $b\sim 100$ and ${\cal R}_6\sim 0.1$ are well conceivable.

\placetable{tab1}

As one can see from Fig.~1, even in the extremal case (2), quite a large number 
of normal pulsars lie below the critical line, meaning that the VG-CR cannot 
form in these objects.  It is important
to notice that two well known drifting subpulse pulsars, PSRs B0820$+$02 and 
B1944$+$17 \citep[][Table 2]{r86} also lie
below the extremal line (2), thus some kind of VG inner accelerator should 
operate in this region. We can see three possible
solutions to this problem: (i) the estimation of ion critical temperature 
(eq.[1]) is inadequate, at least for older pulsars,
(ii) some radio pulsars, especially the older ones, are BPCSS instead of 
neutron stars, as originally proposed by \citet{xqz99,xzq00}, and (iii) these 
pulsars are driven by the VG-ICS inner accelerator. Let us briefly 
discuss these three
possibilities. The possibility (i) is least promising. In fact, even 
if the ion critical temperature $T_i$ is 
underestimated by a factor of 10, one can move the critical line (2) 
down only by a factor of 18, which still leaves 
quite a
number of normal pulsars below it. The possibility (ii) would improve 
the situation radicaly, as the binding energy in bare
polar caps of strange stars is almost infinite for both electrons 
and positrons \citep{xqz99,xzq00}. However, the existence of BPCSS is rather speculative
and certainly not widely accepted. The possibility (iii) seems to be
a viable option. In fact, in the ultra-high magnetic field the pairs 
are produced near the kinematic threshold 
$(\hbar\omega/2mc^2)\cdot h/{\cal R}\sim 1$ \citep{dh83}. 
The near threshold critical lines for the VG-ICS formation are described 
by the equation (\ref{a14}) in the \S~3.
We can use $b=100$ and ${\cal R}_6^{0.8}(k\cdot{\cal F})^{0.7}=0.1$ to obtain 
the extremal critical line $\dot{P}_{-15}= 0.003\cdot P^{-2.2}$, which is 
presented as the line (3) in Fig.~1. Since the dipolar magnetic field in the considered 
region of $P-\dot{P}$ diagram is low ($B_d\sim 10^{11}$~G), it is conceivable 
to increase the value of $b$ to about  
few hundreds (and still not to exceed $B_q\sim 4.4\cdot 10^{13}$~G).
Thus, it is obvious that all normal pulsar can be explained by VG-ICS and/or 
VG-CR inner accelerator, without invoking the
BPCSS conjecture \citep{xqz99,xzq00}. 

As one can see from Fig.~1, the near treshold ICS gap (line 3) makes 
an important difference from the near threshold CR gap (line 2), which follows from 
the fact that in the ICS case the electrons mean free 
path $l_e\sim h$ is important, while in the VG case $l_e\ll h$ is
negligible \citep{zhm00}. Moreover, in the assymptotic case the gap height $h\sim l_e$ 
is quite long (see eq. [\ref{a12}])
and the polar cap heating is considerable. However, since $l_e\propto B_s^{-1}$, 
higher surface magnetic field
$B_s$ leads to much lower gap heights in the near threshold case (see eq. [\ref{a11}]) 
and correspondingly lower surface temperatures $T_s$.
 
\placefigure{fig1}

 Our working hypothesis is therefore that the observed normal radio pulsars are driven 
by VG inner accelerator. This means 
that $T_s/T_i<1$ for pulsars located above the critical lines (1), (2) or (3). 
We further speculate, that shorter or longer episodes
of $T_s/T_i<1$ (when the VG gap cannot form) could be attributed to the well 
known and common phenomenon of pulse nulling.
As \citet{r86} demonstrated, nulling occurs simultaneously in both core and 
conal components of complex profiles,
meaning that this phenomenon is associated with neither core nor conal emission. 
However, the core single pulsar apparently
do not null. This striking property seems to have natural explanation within 
the framework of our model. In fact, the core single pulsars
occupy regions of the $P-\dot{P}$ diagram well above the critical lines 
for which $T_s=T_i$. Taking $P\sim 0.3$~s and 
$\dot{P}_{-15}\sim 30$ as the average values in the group of core single 
pulsars \citep[see Fig.~6 in][]{gs00} we obtain from 
equation~(\ref{tsti}) that $T_s/T_i=0.4$ even for $k\cdot{\cal F}=b={\cal R}_6=1$, so 
the VG-CR gap can always form in these objects. We intend to explore the above mentioned idea in a
separate paper.

 \section{Pulsar death lines and PSR J2144$-$3933}

In this section we examine an influence of strong surface magnetic 
fields required by the ion binding energy problem on the location of death lines 
on the $P-\dot{P}$ diagram. We are specially interested in PSR J2144$-$3933 with the
extremal period $P=8.5$~s, which lies well beyond all conventional death lines.
An approximate condition for pair creation in strong magnetic 
field $B$ was given 
by \citet{e66}. In the limit of high photon energies 
$\hbar\omega\stackrel{>}{\sim}2mc^2$,
the conversion rate depends sensitively on 
the parameter 
$\chi=(\hbar\omega/2mc^2)(h/{\cal R})(B/B_q)\stg 1/15$. 
 \citet{cr93} considered the problem of death lines in 
nondipolar configurations of the surface magnetic field at the pulsar polar 
cap, using the asymptotic approximation described by the above condition
$\chi\stg 1/15$. We first apply their results to the new 8.5 second
period pulsar. 
With very curved lines and strong field the gap height 
is $h\sim(B_d/B_s)^{1/2}R\left(2\pi R/cP
\right)^{1/2}$. When this reduction of the gap height is taken into
account, then the corresponding death line equation takes the form 
$$
\log B_d=1.9\log P-\log B_s+0.6\log{\cal R}+21 ,
 $$
where we introduced the radius of curvature explicitely.
Setting $P=8.5$~s and $B_d=4\times 10^{12}$~G, we find that 
$B_s\stg 10^{13}$~G and ${\cal R}\ll 10^6$~cm. 
 
The above asymptotic considerations suggest that the surface magnetic field in PSR J2144$-$3933 should be
very strong and extremaly curved \citep[see also][]{ymj99}. However,
the asymptotic condition for magnetic pair creation  
is not valid for magnetic fields $B_s\stg 0.1B_c\approx
5\times 10^{12}$~G, which is much lower than surface fields inferred just above, by means of
this approximation. Thus we have to use the near threshold condition $\hbar\omega\cdot\sin\theta\stg 2mc^2$, discussed 
generally in the \S~3. A general near threshold 
CR death line is expressed by equation (\ref{a10}). 
For PSR J2144$-$3933 $P=8.5$~s and $\dot{P}_{-15}=0.475$ and thus 
we obtain a condition ${\cal R}_6^{2}\cdot b^{0.5}=0.13$.
Treating this  
as a general condition we obtain a death 
line $\dot{P}_{-15}=3\cdot 10^{-5}P^{4.5}$, which is presented as
the line (4) in Fig.~1. Since in PSR J2114$-$3933 $B_d=4\cdot 10^{12}$~G
(and $B_s=b\cdot B_d<B_q=7.6\cdot 10^{13}$~G), then 
$2\stl b\stl 20$, which gives $0.3\stg{\cal R}_6\stg 0.17$, respectively.

It is also interesting to compare the near threshold ICS induced death line with the 
CR induced death line represented
by the line (4) in Fig.~1. Using the equation (\ref{a15}) and 
putting $b=100$ and ${\cal R}_6=0.1$ into it, 
one obtains death line
$\dot{P}_{-15}=0.0002\cdot P^{-0.28}$ which is presented as the line (5) in 
Fig.~1. Obviously, this critical death
line can explain all normal (non-millisecond) pulsars.
However, the case of PSR J2114$-$3933 can be interpreted with much less ad hoc field
configuration, for example $b\approx 2$ and ${\cal R}\sim 1$.

\section{Conclusions}

In this paper we have examined an influence of extremely strong and curved 
surface magnetic field on the long
standing binding energy problem \citep[for review see][]{xqz99}. We have 
demonstrated within CR and/or ICS driven
pair production model that formation of a vacuum gap above the pulsar polar
cap is in principle possible, provided that $B_s\gg B_d$ and ${\cal R}\ll R$. 
We have also addresed in this paper a very interesting and currently topical problem of
why so many radio pulsars are beyond the conventional death lines, where 
theoretically they
should not be converting high energy photons into electron-positron pairs. 
PSR J2144$-$3933 is the pulsar of our special interest
with $P=8.5$~s and 
$\dot{P}=4.75\times 10^{-16}$ (thus $B_d=4\times 10^{12}$~G), 
which is located extremely far beyond conventional death lines
(circle in Fig.~1).  Assuming
strong multipolar surface magnetic field suggested by the binding energy 
problem, we can explain this extremal object without 
invoking any different radiation mechanism than that for ordinary
pulsars. A value of $B_s \stg 10^{13}$~G and 
${\cal R}\stl 10^5$~cm seems to fit the constraint imposed by the 
8.5~s period well. \citet{ymj99} have noticed that extremely 
twisted field could marginally explain the pulsar location 
in the $B_d-P$ diagram, but
they consider such fields unlikely. However, they did not consider a sunspot 
like configuration that we use in this paper. 

There is a growing evidence that quite a large number of conal-type 
pulsars \citep{r86} with drifting subpulses (grazing 
cuts of the line-of-sight) or periodic
intensity modulation (central cuts of the line-of-sight) have the RS75-type 
polar gap accelerators with curvature
radiation dominated magnetic pair production \citep{rs75,dr99,vj99,gs00,xqz99}. 
We argue in this paper that in such cases the surface magnetic field 
penetrating the polar gap should be dominated by a strong 
multipolar (presumably sunspot-like magnetic field with 
$B_d\ll B\stl B_q$ and ${\cal R}\ll R$) components, reconnecting with a global 
dipole field well before the radio emission region.
\citet{zhm00} concluded in their recent paper that it is not necessary 
to postulate ad hoc multipolar field configuration to explain 
PSR J2144$-$3933. In fact, they demonstrated that the ICS induced SCLF accelerator
produces death lines that include PSR J2144$-$3993. However, we would like 
to strongly emphasize that the SCLF model is unable to
explain the subpulse drift phenomenon, which seems to be a common phenomenon, at 
least among the conal-type profile pulsars 
\citep{r86,gs00}, although there is no direct evidence for conal type emission in PSR J2144$-$3933.

As can be seen from Fig.~1, the VG formation is in principle possible 
for all normal pulsars (excluding binary and millisecond ones), 
provided that the surface magnetic field is extremaly strong and curved. 
Pulsars above the line (1) can form
the VG-CR inner accelerator even with relatively low surface 
field $B_s\stl 5\cdot B_d$. However, pulsars below this line
require much higher fields $B_s\gg B_d$ to form the VG-CR ($b\sim 100$ at 
line (2) but it has to decrease towards the line (1),
as the actual field should not exceed the critical field $B_c\sim 5\cdot 10^{13}$~G). 
Below the line (2) one cannot form
the VG-CR accelerator unless $B_s\gg 100B_d$. However, it is possible 
to form a VG-ICS inner accelerator with $b\sim 100$ 
and ${\cal R}_6\sim 0.1$ on the line (3). Since the dipolar 
field $B_d\sim 10^{11}$~G around this line, one can use
$b$ even larger than 100, and therefore all normal pulsars lie above critical 
lines for the VG formation. The general
conslusion is that the VG formation, either CR or ICS induced, require strong 
and curved surface magnetic field with radius of
curvature ${\cal R}\sim 10^5$~cm and the field strenght $B_s\gg B_d$, where 
$B_c$ is the photon splitting level.
This implies, that radio pulsars (at least non-milisecond ones) are neutron 
stars with surface magnetic field strenght 10-100 times the surface dipolar component derived
from $P$ and $\dot{P}$ values. Such NS can form 
VG-CR or VG-ICS inner accelerators, which discharges
via a number of localised sparks. Both VG-CR and VG-ICS produce sparks, importance 
of which for the mechanism of coherent pulsar radio emission
was recently emphasised by \citet{xzq00}.

The sparking discharge within a VG inner accelerator
is also a natural explanation of the subpulse drift/periodic modulation. 
The group of pulsars with clearly drifting subpulses seem to favour 
the VG-CR model, although the VG-ICS model
is not excluded. Few well know drifters lying below the line (2) 
in Fig.~1 seem to require the ICS contribution
or even domination of the gap discharge. 
It is important to emphasize that pulsars with signatures of periodic 
subpulse modulations seem to be distributed
all over the bulk of the $P-\dot{P}$ plane \citep{r86}.
Therefore, it seems unlikely that pulsars with drifting subpulses represent 
the VG accelerator while others are driven
by the SCLF accelerator. We suggest that normal radio pulsars are those 
neutron stars which can develope the VG inner accelerator above their polar caps. 
This interpretation is consistent with the current status of theory of the coherent
pulsar radio emission, which also seem to favour the sparking VG model, at least 
in normal pulsars \citep{gs00,mgp00,qlzh00}. The millisecond pulsars
are probably driven by the SCLF inner accelerator. This problem will be examined 
in the forthcoming paper.

We would like to note that a sunspot-like configuration conjectured in this 
paper for VG formation and 
supported by the case of PSR J2144$-$3933 is also suggested by
the spin-down index of PSR B0540$-$69 \citep[CR93,][]{rzc97}. Also \citet{cz99}, analyzing 
the X-ray emissions from regions near
the polar cap of the rotation powered pulsars, 
argued that $B_s\sim 10^{13}$~G and ${\cal R}\sim 10^5$~cm, which agrees well with our
independent estimates. 
As already mentioned, \citet{gs00} attempting to explain morphological 
differences in pulse shapes 
and variations in polarization properties of different profile types as 
well as the subpulse drift rates, concluded that
the surface magnetic field at the polar cap should be dominated by 
sunspot-like structures. 
It seems that there is a growing evidence of such small scale anomalies in surface 
magnetic field of radio pulsars \citep[see also][]{ps96,r91}, and our 
paper gives independent arguments supporting this idea.

Polarization studies in radio pulsars seems to suggest that 
significant part of the radio emission arises from regions
in the magnetosphere where the magnetic field is largely 
dipolar. Also studies concerning the 
morphology of pulse profiles is consistent with dipolar magnetic 
field structure in the radio emission altitudes which is thought
to be arising close to the stellar surface given by the relation
$r\sim 50\cdot R\cdot P^{0.3}$ \citep{kg97,kg98}. 
Thus, if there are strong and complicated surface 
magnetic fields $B_s\sim 100B_d$ on the stellar surface it 
is important to access whether
complicated fields would decay fast enough with altitude
such that in the emission region the magnetic field structure 
is almost purely dipolar as constrained by observations. 
On the other 
hand, the radius of curvature of surface field lines should be 
about $10^{5-6}$~cm, which
means that the structure of the gap magnetic field is determined by 
multipoles higher than the quadrupole. 
Here we consider the model of the magnetic field to be sunspot like near 
the surface of the neutron star while the global dipolar magnetic field is 
that of a star centered dipole. Such small scale magnetic fields on the 
surface of the neutron  star has also been considered by several authors
to explain the radio emission properties from pulsars \citep[e.g.][]{kro91,vr80}.
In the `crustal-model' of the neutron star where the magnetic field is 
generated in neutron star due to the currents in the crust, it is predicted 
that these models are only capable of producing small scale magnetic fields 
\citep{uly86}. These small scale fields can be modelled as small
current loops giving rise to dipolar fields which are oriented 
in different directions all over the stellar surface and superposed on the
global dipole field. Following this one can express the magnetic field 
$B(r)$ at a distance $r$ from the stellar center as the multipolar expansion  
$B(r)\approx B_d(R/r)^3+\Sigma_{l\geq3}B_l(R/r)^{l+1}$, where $l=2,3,4...$
correspond to dipole, quadrupole, octupole etc. and $B_l$ is the 
magnetic field strength of a given pole at the neutron star's surface.
The number of reversals of the magnetic field across the stellar surface
depends on the order of the multipole in question. For example a pure dipolar
field will have 2 reversals while a pure quadrupole will have 4 reversals 
and in general for a multipole of order $l$ there will be $2^{l-1}$ reversals. 
Here we consider the multipolar field which has sun-spot like magnetic 
loops with typical radius of curvature ${\cal R}\sim 0.1R\approx 10^5$~cm. This type of structure
corresponds to multipoles of order $l>4$. The actual multipole order of a crust origin field clump
can be estimated as $l\sim\Delta r/R$, where $\Delta r\sim 0.1R$ is a characteristic crust thickness.
For typical pulsars with $P\sim1$ second the radio emission arises
at altitudes $r\sim 50\cdot R$, and the ratio $B_d/B_{l>4} > 1$ in the emission region even if
$B_l/B_d\sim 100$ at the surface. This means that radio emission arises from dipolar
field lines even if there are strong, small scale multipolar anomalies on the surface
of the polar cap. The field surface strenght of a local multipole is $B_s\sim(R/\Delta R)^mB_d$, where
$m=1$ or 2 depending on relative orientation of adjacent magnetic moments \citep[e.g.][]{a81}. This gives
an estimate $B_s/B_d\sim10\div 100$, as inferred in this paper using the binding energy arguments.

We have shown in this paper that VG formation in typical pulsars requires the crust-origin
surface multipolar field to be much stronger (10-100 times) than the dipolar surface
component of the star centered global field. In order to apply to the millisecond pulsars,
we would have to invoke magnification factors $10^4-10^5$. This would mean that the millisecond pulsars have
surface magnetic fields of the same order as the typical (normal) pulsars. \citet{cgz98}
came to similar conclusion analysing the non-thermal X-ray emission from outer gaps of rotation
powered pulsars. Whether such strong surface magnetic field can exist in millisecond pulsars is not known.
We would like to mention here one constraint found in the literature. \citet{a93} has shown that the location
of the spin-up line in the $P-\dot{P}$ diagram constraints the large scale anomlies of the magnetic field
at the surface to no more than about 40\% of the surface strenght of dipolar field.
This constraint does not concern the crust-origin small scale anomalies invoked in this paper, which will not affect
the dipolar structure of the magnetic field in the radio emission region. However, if such extremely strong fields
are excluded in the millisecond pulsars, then their inner accelerator must be of SCLF type, which implies
the maser kind of radio emission mechanism \citep{kmmu87,kmm92}.

It is worth to emphasize that all our conclusions  presented in this paper concern vacuum gaps in
neutron stars. Therefore, we claim that it may not be necessary to invoke the BPCSS conjecture 
proposed by \citet{xqz99}, according
to which pulsars, at least those with drifting subpulses, are strange stars with bare polar caps (for review see
Xu et al. 2000a). However, we have to admit that we cannot exclude the BPCSS hypothesis.

A complicated magnetic field with radius of curvature much smaller than 
the star radius has to be confined to the neutron star crust.
\citet{mkb99} examined the evolution of multipole components generated by
currents in the outer crust. They found that mostly low order multipoles 
contribute to the required small radii of curvature, and that the
structure of the surface magnetic field is not expected to change 
significantly during the radio pulsar lifetime.
The important question is if the crust can support surface magnetic
fields with a magnitude approaching $10^{14}$~G. Equating the magnetic 
pressure of a strong multipolar field to the crustal stress one concludes 
that the maximum field which the neutron star can sustain is approximately 
about $10^{14} - 10^{15}$ G \citep{td95}. Thus, our inferred 
magnitudes $B_s>10^{13}$ G are in the regime 
where cracking of the neutron star crust will still not occur.

We would like to point out a small weakness of our paper, namely an apparent lack of pulsars 
in a valley between the marginal line (4) in Fig.~1 (including the extremal pulsar PSR J2144$-$3933) and the
right-hand boundary of the bulk of $P-\dot{P}$ distribution. The lack of pulsars at the 
long period regime near the
8.5 second pulsar is understandable, since detectibility of long period pulsars is 
much smaller than that
of shorter ones, as pointed out by \citet{ymj99}. The lack of shorter period pulsars 
within this valley has been explained by
\citet{zhm00}, who invoked a death line below which pair production is not supported by the
SCLF-ICS mechanism (line IV in their Fig.~1). Within 
our VG scenario, this might
just mean that strong multipole fields are possible but not common in old pulsars. 
Another possible explanation concerns
a luminosity issue. Let us notice that PSR J2144$-$3933 is a very old (281 Myr), extremaly weak (4 mJy) 
and close to the Earth (0.19 kpc)
pulsar. It would not probably be detected if it was located just a bit further away. So 
perhaps the radio 
luminosity of pulsars drops below a detection threshold of a typical 
survey before their inner accelerators stop
completely producing a pair plasma.

Finally, we would like to emphasize once again that vacuum gaps which we 
have shown to exist in pulsars,
produce sparking discharges. These isolated sparks seem to be naturally 
involved with drifting subpulses observed in typical pulsars
\citep[e.g.][]{r86,dr99,gs00}. On the other hand, spark-associated models of the 
coherent pulsar radio emission
\citep{qlzh00,mgp00} critically depend on existence of non-stationary vacuum gaps, and 
therefore our paper supports these ideas on fundamental grounds.
 
\begin{acknowledgements}
This paper is supported in part by 
the Grant 2~P03D~008~19 of the Polish State Committee for Scientific
Research. We are indebted to the referee B. Zhang for insightful comments 
and constructive criticism which greatly
helped to improve the final version of the paper. We are also thankful to 
K.S. Cheng, J. Kijak, D. Khechinashvili, G.~Machabeli, 
G. Melikidze, B. Stappers and B. Rudak for helpful comments 
and discussions.
We also thank the Max-Planck-Institut f{\"u}r Radioastronomie and 
the Astronomical Institute ``Anton Pannekoek'' University of Amsterdam
for kind hospitality, where part of this work was done.
DM thanks J. Kepler Astronomical Center for support and 
hospitality during his visit to the institute, where this work was started.
We thank E. Gil for technical assistance.
\end{acknowledgements}

 \begin{deluxetable}{llllll}
\footnotesize
\tablecaption{Four pulsars with clearly drifting subpulses \label{tab1}}
\tablewidth{0pt}
\tablehead{
{PSR} & P & {$\dot{P}_{15}$} & {$L_{x}/10^{30}$}& {$T_s$}  & {$T_i$}\\ 
 & (s)& &(erg/s) &($10^6$K)&($10^6$K) }
\startdata
0031$-$07 & $0.94$ & $0.4$ & $0.012$ & $2.2$ & $2.2$ \\ 
0943$+$10 & $1.1$ & $3.5$ & $0.027$ & $2.8$ &$5.0$ \\ 
2303$+$30 & $1.57$ & $2.9$ & $0.014$ & $2.4$ &  $5.6$ \\ 
2319$+$60 & $2.26$ & $7.0$ & $0.013$ & $2.3$& $8.7$ 
\enddata
\end{deluxetable}

\figcaption[]{The $P-\dot{P}$ diagram for 538 pulsars from the Pulsar
Catalog \citep{tml93}  with measured $\dot{P}$ values.  Three solid
lines are the critical lines for vacuum gap formation corresponding
to different acceleration region models: (1) assymptotic VG-CR
obtained from equation (\ref{dotp}), (2) near threshold VG-CR obtained
from equation (\ref{a9}) for $b=100$ and $(k\cdot{\cal F})^{1.15}\cdot
{\cal R}^{0.64}_6=0.15$, and (3) near threshold VG-ICS models obtained
from equation  (\ref{a14}) for $b=100$ and $(k\cdot{\cal
F})^{0.7}\cdot{\cal R}^{0.8}_{6}=0.1$ models, respectively.  Two
dashed lines represent near threshold death lines: (4) obtained for CR
from equation (\ref{a10}) for ${\cal R}^2_6\cdot b^{0.5}=0.13$ and (5)
obtained for ICS from equation (\ref{a15}) for $b=100$ and ${\cal
R}_6=0.1$, respectively. Three dotted lines  represent a constant
dipole magnetic field $B_d=10^{11},10^{12}$ and $10^{13}$~Gauss,
respectively.  Pulsars with drifting subpulses are marked  by crossed
circles and PSR J2144$-$3933 is marked by open circle.\label{fig1}}

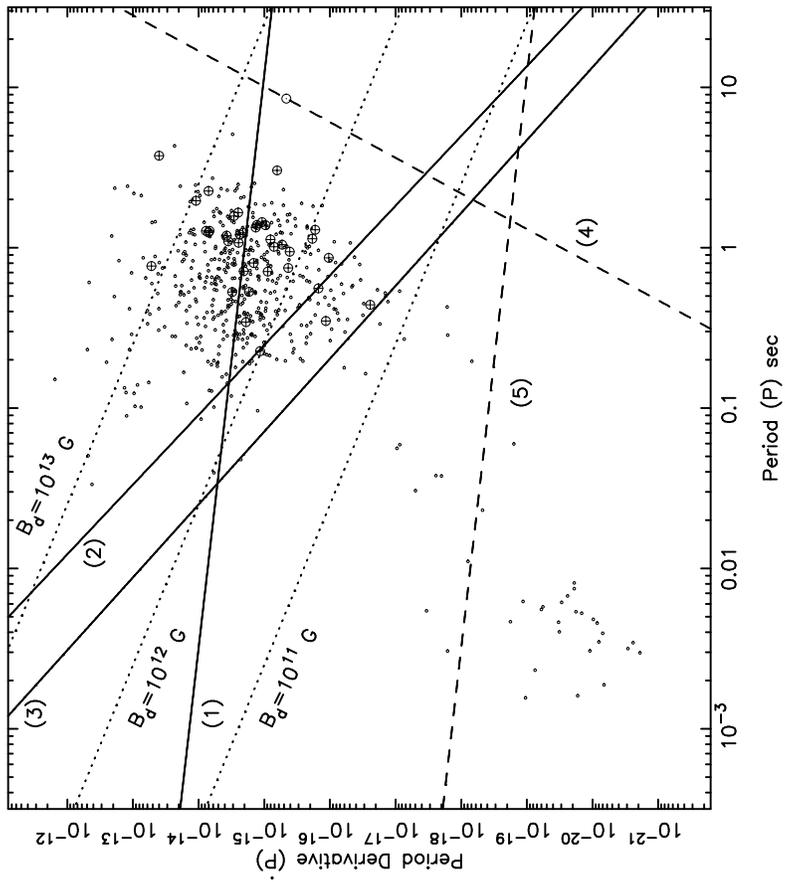
\end{document}